\documentclass[preprint,aps,showpacs,nofootinbib,preprintnumbers,amsmath,amssymb]{revtex4-1}
\usepackage{}
\usepackage{epsfig}
\usepackage{subfigure}
\usepackage{dcolumn}% Align table columns on decimal point
\usepackage{bm}% bold math
\usepackage[usenames ,dvipsnames]{xcolor}
\usepackage{slashed}

\usepackage{graphicx,color}

\begin{document}
\title{A global $SU(3)/U(3)$ flavor symmetry analysis \\ for $B\to PP$ with $\eta-\eta'$ Mixing}

\author{Yu-Kuo Hsiao$^{1,2,3}$, Chia-Feng Chang$^4$ and Xiao-Gang He$^{5,4,2} $}
\affiliation{
$^1$Chongqing University of Post \& Telecommnications, Chongqing, 400065, China\\
$^{2}$Physics Division, National Center for Theoretical Sciences, Hsinchu, Taiwan 300\\
$^{3}$Department of Physics, National Tsing Hua University, Hsinchu, Taiwan 300\\
$^4$Department of Physics, National Taiwan University, Taipei, Taiwan 107\\
$^5$Department of Physics and Astronomy, Shanghai Jiao Tong University, Shanghai, China}

\begin{abstract}
A large number of new experimental data on $B$ decay into two light pesudoscalar ($P$) mesons  have been collected by the LHCb collaboration. Besides confirming information on $B_{u,d} \to PP$ decays obtained earlier by B-factories at KEK and SLAC, new information on $B_s\to PP$ and also more decay modes with $P$ being $\eta$ or $\eta'$ have been obtained. Using these new data, we perform a global fit for $B \to PP$ to determine decay amplitudes in the framework of $SU(3)/U(3)$ flavor symmetry. We find that $SU(3)$ flavor symmetry can explain data well. The annihilation amplitudes are found to be small as expected. Several CP violating relations predicted by $SU(3)$ flavor symmetry are in good agreement with data. Current available data can give constraints on the amplitudes which induce $P = \eta,\;\eta'$ decays in the framework of $U(3)$ flavor symmetry, and can also determine the $\eta-\eta'$ mixing angle $\theta$ with $\theta = (-18.4\pm1.2)^\circ$. Several $B \to PP$ decay modes which have not been measured are predicted with branching ratios accessible at the LHCb. These decays can provide further tests for the framework of $SU(3)/U(3)$ flavor symmetry for $B$ decays. 
\end{abstract}
%=========================================================================
\maketitle

%=========================================================================
\section{introduction}
A large number of experimental data on $B$ decay into two pesudoscalar ($P$) mesons have been collected by the LHCb collaboration. Besides confirming information on $B_{u,d}\to PP$ obtained earlier by B-factories at KEK and SLAC, new information on $B_s$ decays have been obtained~\cite{pdg,average}.  The new information can provide more insight about interactions responsible for $B$ decays. $B \to PP$ are induced at one loop level in the standard model (SM).
These decay modes being rare ones are expected to be sensitive to new physics beyond the SM.  Before claiming the existence of any new physics beyond it is necessary to have the SM interactions be well understood.  $B\to PP$ decays have been studied extensively in different ways. The main methods are QCD based perturbative calculations~\cite{QCDF2001,pQCD2001,SCET2001}
and $SU(3)$ flavor symmetry~\cite{Zeppendfeld,Chau,Chau:1990ay,Dighe:1997hm,Gronau,
Chiang:2004nm,Deshpande:1994ii,SU3_anni1,Savage:1989ub}. 

The $SU(3)$ flavor symmetry approach has the advantage of being detailed dynamics independent. The decays are described by several $SU(3)$ invariant amplitudes which can lead to relations between different decay modes, but this approach by it-self cannot determine the size of the amplitudes. The QCD based perturbative approach being dynamic models, for example, the QCD factorization (QCDF)~\cite{QCDF2001}, 
perturbative QCD (pQCD)~\cite{pQCD2001}, 
and soft-collinear effective theory (SCET)~\cite{SCET2001}, can calculate the very precisely measured CP violation asymmetry  
${\cal A}_{CP}(\bar B^0\to \pi^+ K^-)=(-8.2\pm 0.6)\%$~\cite{pdg,average} for $\bar B^0\to \pi^+ K^-$ decay. If the theory is universally valid they should be able to make accurate predictions for CP violation in other 
$B\to PP$ decays. These methods, however, all predict ${\cal A}_{CP}(\bar B^0\to \pi^+ K^-)\approx {\cal A}_{CP}(B^- \to \pi^0 K^-)$, which is in contradiction with experimental observation. Therefore ${\cal A}_{CP}(\bar B^0\to \pi^+ K^-)\ne {\cal A}_{CP}(B^- \to \pi^0 K^-)$ challenges these theories~\cite{Hou:2006jy,Li:2009wba,Cheng:2011qh}.
On the other hand, the analysis based on the $SU(3)$ flavor symmetry can be 
advantageous, where the different decay modes can be related and
the relevant decay amplitudes be extracted from the data, despite of their unclear sources. A consistent solution for these CP violating asymmetries can be found. When sufficient data become available, 
the $SU(3)$ invariant amplitudes can be fixed and predictions 
be made,
and the theory 
be tested.
$SU(3)$ analysis may play a role to bridge dynamic theory and experimental data to provide some understanding of SM predictions for $B$ decays.

The $SU(3)$ flavor symmetry has been wildly used for the studies in the SM 
for two-body and three-body mesonic $B$ decays~\cite{SU3_eta,SU3_PPP},
the extraction of the weak phase~\cite{SU3_gamma1,SU3_gamma2}, and the constraint on new physics~\cite{SU3_np}.
In its extended version, the two-body anti-triplet $b$-baryon decays of 
${\cal B}_b\to {\cal B}_n M$ and ${\cal B}_b\to {\cal P}_c M$ decays
can be studied~\cite{SU3_Lb1,SU3_Lb2,SU3_Lb3}, where ${\cal B}_n$ and ${\cal P}_c$ 
stand for the baryon and pentaquark state, respectively, with $M$ the recoiled meson. 
To make sure $SU(3)$ flavor symmetry framework is valid for $B$ decays, experimental test should be performed.
Due to the fact that the Belle and Babar detectors at B-factories can only study 
$B_u$ and $B_d$, but not $B_s$ decays, the $SU(3)$ flavor symmetry have not been well tested. With the running of LHC, the LHCb has been able to obtain valuable data not only on $B_{u,d}$, but also $B_s$ decays, one can therefore test more thoroughly the $SU(3)$ flavor symmetry for $B \to PP$ decays~\cite{He:2013vta}. 
When more $b$-baryon decays are measured, $SU(3)$ can also be tested for the $b$-baryon sector. 
Experimentally, the data collections for the $B\to PP$ decays are in fact still not satisfactory. 
For example, $\bar B_s^0 \to K^0 \pi^0$ and $\bar B_s^0\to K^0 \bar K^0$ 
%and 
and $\bar B_s^0\to \eta\eta,\eta\eta'$ have not been observed yet.
Some decays with small branching ratios expected from theoretical considerations, such as those 
decays, $\bar B^0\to K^+ K^-$, $\bar B_s^0 \to \pi^+ \pi^-$, 
and $\bar B_s^0 \to \pi^0 \pi^0$ dominated by the annihilation contributions~\cite{SU3_anni1,SU3_anni2} 
need further confirmation from data. Taking this positively, one can then use $SU(3)$ flavor symmetry framework to predict
their branching ratios as further tests.

In this work, we will perform an updated global analysis for $B \to PP$ using the latest experimental data 
based on flavor symmetry. Without including $\eta$ and $\eta'$ in the final states, $SU(3)$ flavor symmetry is sufficient for the analysis.
In order to include them also in the analysis, one needs to modify the analysis method. To this end we will enlarge the symmetry to 
$U(3)$ flavor symmetry, and also to take into account  $\eta-\eta'$ mixing effect to study final states with $P$ being $\eta$ or $\eta'$. 
We find that 
%that 
$SU(3)$ flavor symmetry can explain data well without $P$ being $\eta$ or $\eta'$. The annihilation amplitudes are found to be small consistent with expectations. Several CP violating relations predicted by $SU(3)$ flavor symmetry are found in good agreement with data. Current available data can give constraints on the amplitudes which induce $P = \eta,\;\eta'$ decays in the framework of $U(3)$ flavor symmetry, and the $\eta-\eta'$ mixing angle $\theta$ can also be determined with $\theta = (-18.4\pm1.2)^\circ$ which is consistent with the value given by Particle Data Group from other fittings~\cite{pdg}. Several $B \to PP$ decay modes which have not been measured are predicted with branching ratios accessible at the LHCb. These decays can provide further tests for the framework of $SU(3)/U(3)$ flavor symmetry for $B$ decays. In the following sections, we provide more details of our analysis.

\section{$SU(3)$ decay amplitudes for $B\to PP$}

The quark level effective Hamiltonian responsible
for charmless $B\to PP$ decays can be written as~\cite{buras}
\begin{eqnarray}
H_{eff}^q = {4G_{F} \over \sqrt{2}} [V_{ub}V^{*}_{uq} (c_1 O_1 +c_2 O_2)
-\sum_{i=3}^{11}(V_{ub}V^{*}_{uq}c_{i}^{uc} +V_{tb}V_{tq}^*c_i^{tc})O_{i}],
\end{eqnarray}
with the superscript $q=d(s)$ for $\Delta S=0\,(-1)$ decay modes and
$V_{ij}$ the KM matrix elements.
The coefficients $c_{1,2}$ and $c_i^{jk}=c^j_i-c^k_i$ are the Wilson Coefficients
which have been evaluated by several groups~\cite{buras} with $|c_{1,2}|>> |c_i^{jk}|$.
The operators $O_i$  that consist of quarks and gluons can be written as
\begin{eqnarray}
\begin{array}{ll}
O_1=(\bar q_i u_j)_{V-A}(\bar u_i b_j)_{V-A}\;, &
O_2=(\bar q u)_{V-A}(\bar u b)_{V-A}\;,\\
O_{3,5}=(\bar q b)_{V-A} \sum _{q'} (\bar q' q')_{V \mp A}\;,&
O_{4,6}=(\bar q_i b_j)_{V-A} \sum _{q'} (\bar q'_j q'_i)_{V \mp A}\;,\\
O_{7,9}={ 3 \over 2} (\bar q b)_{V-A} \sum _{q'} e_{q'} (\bar q' q')_{V \pm A}\;,\hspace{0.3in} &
O_{8,10}={ 3 \over 2} (\bar q_i b_j)_{V-A} \sum _{q'} e_{q'} (\bar q'_j q'_i)_{V \pm A}\;,\\
O_{11}={g_s\over 16\pi^2}\bar q \sigma_{\mu\nu} G^{\mu\nu} (1+\gamma_5)b\;,&
O_{12}={Q_b e\over 16\pi^2}\bar q \sigma_{\mu\nu} F^{\mu\nu} (1+\gamma_5)b\;.
\end{array}
\end{eqnarray}
Under $SU(3)$ flavor symmetry, while the Lorentz-Dirac structure and color index are  both omitted,
$O_{1,2}$, $O_{3-6,11}$, and $O_{7-10}$ transform as $\bar 3 + \bar 3' +6 + \overline {15}$,
$\bar 3$, and $\bar 3 + \bar 3'+6 + \overline {15}$, 
respectively~\cite{SU3_anni1,Savage:1989ub,Dighe:1997hm,Zeppendfeld,Chau,Gronau}.
As a result, $H_{eff}^q$ can be decomposed as the matrices of $ H(\overline{3})$,
$H(6)$, and $H(\overline{15})$ with their non-zero entries to be~\cite{SU3_anni1}
\begin{eqnarray}
&&H(\overline{3})^{2}=1\;,\; H(6)^{12}_{1}=H(6)^{23}_{3}=1\;,
\;H(6)^{21}_{1}=H(6)^{32}_{3}=-1\;,\nonumber\\
&&H(\overline{15})^{12}_{1}=H(\overline{15})^{21}_{1}=3,\;H(\overline{15})^{22}_{2}=-2\;,
\;H(\overline{15})^{32}_{3}=H(\overline{15})^{23}_{3}=-1\;,
\end{eqnarray}
for $\Delta S=0$, and 
\begin{eqnarray}
&&H(\overline{3})^{3}=1\;,\; H(6)^{13}_{1}=H(6)^{32}_{2}=1\;,
\;H(6)^{31}_{1}=H(6)^{23}_{2}=-1\;,\nonumber\\
&&H(\overline{15})^{13}_{1}=H(\overline{15})^{31}_{1}=3\;,\;H(\overline{15})^{33}_{3}=-2\;,
\;H(\overline{15})^{32}_{2}=H(\overline{15})^{23}_{2}=-1\;,
\end{eqnarray}
for $\Delta S=-1$. Accordingly,
the $B$ mesons are presented as $B_i=(B_u,  B_d,  B_s) = (B^-, \bar B^0, \bar B^0_s)$,
and for the final state $P$ %to be $\pi,K,\eta_{1,8}$
as the octet of $SU(3)$ representation $M^i_j$ is given by
\begin{eqnarray}
(M_i^j)
&=&
\left(\begin{array}{ccc}
\frac{\pi^0}{\sqrt{2}}+ \frac{\eta_8}{\sqrt{6}} &\pi^+& K^+  \\
\pi^-&-\frac{\pi^0}{\sqrt{2}}+ \frac{\eta_8}{\sqrt{6}}  &K^0  \\
K^- & \bar K^0 &    -2 \frac{\eta_8}{\sqrt{6}}
\end{array} \right)\;,
\nonumber
\end{eqnarray}
along with $\eta_1$ as the singlet of $SU(3)$ to be $(M_{\eta_1})_i^j=\delta_i^j\eta_1$. 
Note that $\bar M=M+M_{\eta_1}/\sqrt 3$ form a nonet of $U(3)$. 
Consequently, 
without appealing to the dynamics of perturbative QCD,
the $B\to PP$ decay amplitudes are given by
\begin{eqnarray}\label{TandP}
A(B\to PP) = <PP|H_{eff}^q|B> = {G_F\over \sqrt{2}}[V_{ub}V^*_{uq} T
+ V_{tb}V^*_{tq}P],
\end{eqnarray}
where the tree amplitude  $T$ for $B \to PP$ can be parametrized by $SU(3)$ invariant amplitudes.
If one wants to include $\eta_1$ and $\eta_8$ into consideration, one may want to enlarge the analysis with $U(3)$ flavor symmetry. The $SU(3)/U(3)$ invariant amplitudes are given below~\footnote{
By treating $\eta_1$ as a $SU(3)$ singlet, we can form
another $T$ amplitude with $T=T_{\eta_8}+T_{\eta_1}$.
Note that
$T_{\eta_8}$ can be given by using $T$ in Eq.~(\ref{Tamp}) where
$\bar M=M+M_{\eta_1}/\sqrt 3$ is replaced by $\bar M=M$, 
while $T_{\eta_1}$ can be written as~\cite{SU3_eta}
\begin{eqnarray}\label{abcd}
T_{\eta_1} &=& a^T B_i H(\bar 3)^i \eta_1\eta_1 +
b^T B_i M^i_j H(\bar 3)^j \eta_1 + c^TB_i H(6)^{ik}_lM^l_k \eta_1
+ d^T B_i H(\overline {15})^{ik}_l M^l_k\eta_1.\nonumber
\end{eqnarray}
The $a^i$, $b^i$, $c^i$, $d^i$ and $D^i$, $B^i$ amplitudes are related.}
\begin{eqnarray}\label{Tamp}
T&=& A_{\bar 3}^TB_i H(\bar 3)^i (\bar M_l^k \bar M_k^l)
+ C^T_{\bar 3}B_i \bar M^i_k \bar M^k_j H(\bar 3)^j \nonumber\\
&+&\tilde A^T_{6}B_i H(6)^{ij}_k \bar M^l_j \bar M^k_l
+ \tilde C^T_{6}B_i \bar M^i_j H(6)^{jk}_l \bar M^l_k\nonumber\\
&+&A^T_{\overline{15}}B_i H(\overline{15})^{ij}_k \bar M^l_j \bar M^k_l
+C^T_{\overline{15}}B_i \bar M^i_j H(\overline{15} )^{jk}_l \bar M^l_k\nonumber\\
&+&B^T_{\bar 3}B_i H(\bar 3)^i \bar M^j_j \bar M^k_k
+\tilde B^T_{6}B_i H(6)^{ij}_k \bar M^k_j \bar M^l_l\nonumber\\
&+&B^T_{\overline{15}}B_i H(\overline{15})^{ij}_k \bar M^k_j \bar M^l_l
+D^T_{\bar 3}B_i \bar M^i_jH(\bar 3)^j  \bar M^l_l\;,
\end{eqnarray}
with $\tilde C_6$ and $\tilde A_6$ rearranged to be $C_6=\tilde C_6 - \tilde A_6$~\cite{SU3_anni1,
Savage:1989ub,Dighe:1997hm,Zeppendfeld,Chau,Gronau}. 
Expanding the $T$ expressions in Eq.~(\ref{Tamp}),
we obtain the tree amplitudes $T$ in terms of the symmetry invariant amplitudes
without $\eta_8$ and $\eta_1$ in the final states in Table~\ref{su3tb},
while those with $\eta_8$ or/and $\eta_1$ in the final states are given in Table~\ref{su3tb2}.
Note that
the penguin amplitude $P$ can be given by the replacement of the notation of $T$ by $P$
in the $T$ amplitude, such that the hadronic parameters can be
$C^{P}_{\bar 3,6,\overline{15}}$, $A^{P}_{\bar 3,\overline{15}}$,
$B^{P}_{\bar 3,6,\overline{15}}$, and $D^{P}_{\bar 3}$.

The dynamics of the interactions
are all lumped into the invariant amplitudes, one cannot calculated the values for $A_i$, $B_i$
$C_i$, and $D_i$ just from symmetry considerations, 
and therefore
in our later analysis we will reply on experimental data to
determine them. Note that $A_i^{T,P}$, $B_i^{T,P}$
are referred as annihilation amplitudes because the $B$ mesons are first
annihilated by the interaction Hamiltonian and two light mesons are then created and are expected to be smaller than $C_i$ and $D_i$ amplitudes. 

Further simplification can be made because the operators for the tree and penguin contributions for the representations of $6$ and $\overline {15}$, have the same structure, the differences for related amplitudes are caused by differences of the Wilson Coefficients (WC) in the Hamiltonian. Using WC obtained in Ref.~\cite{buras}, we use the numerical relations obtained in Ref.~\cite{SU3_eta}, 
$C^P_6(B^P_6)=-0.013C^T_6(B^T_6)$, and $C^P_{\overline {15}}(A^P_{\overline {15}},B^P_{\overline {15}})+0.015 C^T_{\overline {15}} (A^T_{\overline {15}},B^T_{\overline {15}})$, respectively.
We comment that in finite order perturbative calculations
the above relations are renormalization scheme and scale dependent.
One should use a renormalization scheme consistently. We
have checked with different renormalization schemes and find that numerically
the changes are less than 15\% for different schemes. In our later analysis, we will use the above relation.
Moreover, since an overall phase can be removed without loss of generality,
by setting $C^P_{\bar 3}$ to be real, there can be totally 25 real independent
parameters for $B\to PP$ in the SM with $SU(3)/U(3)$ flavor symmetry, 
given by\\[2 mm]
\hspace*{2 cm}
$C_{\bar 3}^P,\;\;C_{\bar 3}^T e^{i\delta_{\bar 3}},\;\;
C^{T}_6e^{i\delta_6},\;\;
C^{T}_{\overline{15}}e^{i\delta_{\overline{15}}},\;\;
A^T_{\bar 3}e^{i\delta_{A^T_{\bar 3}}},\;\;
A^P_{\bar 3} e^{i\delta_{A^P_{\bar 3}}},\;\;
A^T_{\overline{15}} e^{i\delta_{A^T_{\overline{15}}}},\;\;\\
\hspace*{2 cm}
B^{T}_{\bar3}e^{i\delta_{B^T_{\bar 3}}},\;\;
B^{P}_{\bar3}e^{i\delta_{B^P_{\bar 3}}},\;\;
B^{T}_{\bar6}e^{i\delta_{B^T_{\bar 6}}},\;\;
B^{T}_{\overline{15}}e^{i\delta_{B^T_{\overline{15}}}},\;\;
D^{T}_{\bar3}e^{i\delta_{D^T_{\bar 3}}},\;\;
D^{P}_{\bar3}e^{i\delta_{D^P_{\bar 3}}}.$\\[2mm]
To obtain the amplitudes for $B$ decays with at least one $\eta (\eta')$
in the final states, one also needs to consider $\eta_1 - \eta_8$ mixing,
\begin{eqnarray}
\left(\begin{array}{c}
\eta\\
\eta'
\end{array}\right)
=
\left(\begin{array}{cc}
cos\theta&-sin\theta\\
sin\theta&cos\theta
\end{array}\right)
\left(\begin{array}{c}
\eta_8\\
\eta_1
\end{array}\right)\;,
\end{eqnarray}
where $\theta$ can be determined by fitting $B\to PP$ data.

%===========================
\begin{table}[t]
\caption{Decay amplitudes for $B\to PP$ without $\eta_8$ and $\eta_1$.\label{su3tb} } 
\footnotesize
\begin{eqnarray}
\begin{array}{ll}
\Delta S = 0 &\Delta S = -1\\
T^{B_u}_{\pi^-\pi^0}(d) = {8\over \sqrt{2}}C^T_{\overline {15}},
&T^{B_u}_{\pi^0K^-}(s)= {1\over \sqrt{2}} (C^T_{\bar 3}
  - C^T_{6} + 3A^T_{\overline {15} } +7 C^T_{\overline {15}
  })\;,\\
T^{B_u}_{K^- K^0}(d)=
C^T_{\bar 3} - C^T_6 + 3 A^T_{\overline {15}} -C^T_{\overline {15}},
&T^{B_u}_{\pi^-\bar K^0}(s)= C^T_{\bar 3}
 - C^T_{6} + 3A^T_{\overline {15}} -  C^T_{\overline {15}},\\
T^{B_d}_{\pi^+\pi^-}(d) = 2A^T_{\bar 3} +C^T_{\bar 3}
 + C^T_{6} + A^T_{\overline {15} } + 3 C^T_{\overline {15}},
&T^{B_s}_{K^+K^-}(s)= 2A^T_{\bar 3} +C^T_{\bar 3}
+ C^T_{6} + A^T_{\overline {15} } + 3 C^T_{\overline {15}},\\
T^{B_d}_{K^- K^+}(d)= 2(A^T_{\bar 3}  +  A^T_{\overline {15}}),
&T^{B_s}_{\pi^0\pi^0}(s) = \sqrt{2}(A^T_{\bar 3}+ A^T_{\overline {15}}),\\\
T^{B_d}_{\pi^0\pi^0}(d)= {1\over \sqrt{2}} (2A^T_{\bar 3} +C^T_{\bar 3}
 + C^T_{6} + A^T_{\overline {15} } -5 C^T_{\overline {15}}),
&T^{B_s}_{K^0\bar K^0}(s)= 2A^T_{\bar 3} +C^T_{\bar 3}
- C^T_{6} -3 A^T_{\overline {15} } - C^T_{\overline {15}},\\
T^{B_d}_{\bar K^0 K^0}(d)= 2A^T_{\bar 3} +
C^T_{\bar 3} - C^T_6 - 3 A^T_{\overline {15}} - C^T_{\overline {15}},
&
T^{B_s}_{\pi^+\pi^-}(s) = 2(A^T_{\bar 3}
+ A^T_{\overline {15}}),\\
T^{B_s}_{ K^0 \pi^0}(d) =
-{1\over \sqrt{2}}(C^T_{\bar 3} + C^T_6 -  A^T_{\overline {15}}
-5 C^T_{\overline {15}}),
&T^{B_d}_{\pi^0\bar K^0}(s)= -{1\over \sqrt{2}} (C^T_{\bar 3}
 + C^T_{6} - A^T_{\overline {15} } -5 C^T_{\overline {15} }),\\
T^{B_s}_{ K^+ \pi^-}(d) =
C^T_{\bar 3} + C^T_6 -  A^T_{\overline {15}} +3 C^T_{\overline {15}},
&T^{B_d}_{\pi^+ K^-}(s) =  
C^T_{\bar 3} + C^T_{6} - A^T_{\overline {15}} + 3 C^T_{\overline {15}}\,.\\\nonumber
\end{array}
\end{eqnarray}
\end{table}

\begin{table}[t]
\caption{Decay amplitudes for $B\to PP$ with at least one of the
$P$ being a $\eta_8$ or $\eta_1$.\label{su3tb2} } \footnotesize
\begin{eqnarray}
\begin{array}{l}
\hspace{-3mm} \left.
\begin{array}{l}
\Delta S = 0\\

T^{B_u}_{\pi^- \eta_8}(d)={2\over \sqrt{6}}
(C^T_{\bar 3}  - C^T_6 + 3 A^T_{\overline {15}} + 3C^T_{\overline
{15}}),\\

T^{B_d}_{\pi^0 \eta_8}(d)= {1\over \sqrt{3}} (-C^T_{\bar
3}  + C^T_6 + 5 A^T_{\overline {15}} + C^T_{\overline {15}}),\\

T^{B_d}_{\eta_8 \eta_8}(d)={1\over \sqrt{2}} (2A^T_{\bar 3} +
{1\over 3} C^T_{\bar 3} - C^T_6 -A^T_{\overline {15}} +
C^T_{\overline {15}}),\\

T^{B_s}_{K^0 \eta_8}(d)= -{1\over
\sqrt{6}}(C^T_{\bar 3}  + C^T_6 -  A^T_{\overline {15}} -5
C^T_{\overline {15}}),\\

T^{B_u}_{\pi^- \eta_1}(d)={{1}\over
\sqrt{3}} (2 C^T_{\bar 3}  + C^T_6 + 6 A^T_{\overline {15}} +
3C^T_{\overline
{15}}\\\;\;\;\;\;\;\;\;\;\;\;\;\;\;\;\;\;\;\;\;\;\; + 3 B^T_6+ 9
B^T_{\overline {15}}+3 D^T_{\bar 3}),\\

T^{B_d}_{\pi^0
\eta_1}(d)={{-1}\over \sqrt{6}} (2 C^T_{\bar 3}  + C^T_6 -10
A^T_{\overline {15}} -5 C^T_{\overline {15}}\\
\;\;\;\;\;\;\;\;\;\;\;\;\;\;\;\;\;\;\;\;\;\; + 3 B^T_6- 15
B^T_{\overline {15}}+3 D^T_{\bar 3}),\\

T^{B_d}_{\eta_1
\eta_8}(d)={1\over 3\sqrt{2}} (2 C^T_{\bar 3}  -3 C^T_6 + 6
A^T_{\overline {15}} + 3C^T_{\overline {15}}\\
\;\;\;\;\;\;\;\;\;\;\;\;\;\;\;\;\;\;\;\;\;\; - 9 B^T_6 + 9
B^T_{\overline {15}}+3 D^T_{\bar 3}),\\
T^{B_d}_{\eta_1
\eta_1}(d)={\sqrt{2}\over 3} (3 A^T_{\bar 3} + C^T_{\bar 3} + 9
B^T_3 +3 D^T_{\bar 3}),\\
T^{B_s}_{K^0 \eta_1}(d)={1\over
\sqrt{3}} (2 C^T_{\bar 3}  - C^T_6 - 2 A^T_{\overline {15}} -
C^T_{\overline {15}}\\
\;\;\;\;\;\;\;\;\;\;\;\;\;\;\;\;\;\;\;\;\;\; - 3 B^T_6 - 3
B^T_{\overline {15}}+3 D^T_{\bar 3}),\\
\end{array}
\right. \left.
\begin{array}{l}
\Delta S = -1\\

T^{B_u}_{\eta_8K^-}(s)=
{1\over\sqrt{6}}(-C^T_{\bar 3}
   + C^T_{6} - 3A^T_{\overline {15}} +9 C^T_{\overline {15}}),\\

T^{B_d}_{\eta_8 \bar K^0}(s)=  -{1\over \sqrt{6}} (C^T_{\bar 3} +
C^T_{6} - A^T_{\overline {15} } -5 C^T_{\overline {15} }),\\

T^{B_s}_{\pi^0\eta_8}(s)= {2\over \sqrt{3}}( C^T_{6} +2
A^T_{\overline {15}} - 2C^T_{\overline {15}}),\\

T^{B_s}_{\eta_8\eta_8}(s)= \sqrt{2}(A^T_{\bar 3} +{2\over 3}
C^T_{\bar 3} - A^T_{\overline {15} } - 2C^T_{\overline {15}})\;,\\

T^{B_d}_{K^- \eta_1}(s)={1\over \sqrt{3}} (2 C^T_{\bar 3}  + C^T_6
+ 6 A^T_{\overline {15}} + 3  C^T_{\overline {15}}\\
\;\;\;\;\;\;\;\;\;\;\;\;\;\;\;\;\;\;\;\;\;\; + 3 B^T_6+{9}
B^T_{\overline {15}}+3 D^T_{\bar 3}),\\

 T^{B_d}_{\bar K^0
\eta_1}(s)={1\over \sqrt{3}} (2 C^T_{\bar 3}  - C^T_6 -2
A^T_{\overline {15}} -  C^T_{\overline {15}}\\
\;\;\;\;\;\;\;\;\;\;\;\;\;\;\;\;\;\;\;\;\;\; - 3 B^T_6 -3
B^T_{\overline {15}}+3 D^T_{\bar 3}),\\

T^{B_s}_{\pi^0
\eta_1}(s)={{-2}\over \sqrt{6}} ( C^T_6 -4 A^T_{\overline {15}} - 2
C^T_{\overline {15}}\\
\;\;\;\;\;\;\;\;\;\;\;\;\;\;\;\;\;\;\;\;\;\; +3 B^T_6 -6
B^T_{\overline {15}} ),\\

T^{B_s}_{\eta_1
\eta_8}(s)={-\sqrt{2}\over 3} (2 C^T_{\bar 3} -6 A^T_{\overline
{15}} - 3  C^T_{\overline {15}}\\
\;\;\;\;\;\;\;\;\;\;\;\;\;\;\;\;\;\;\;\;\;\; - 9 B^T_{\overline
{15}}+3 D^T_{\bar 3}),\\ T^{B_s}_{\eta_1 \eta_1}(s)={\sqrt{2}\over
3} (3 A^T_{\bar 3}+C^T_{\bar 3} + 9 B^T_3+ 3 D^T_{\bar 3})\,.\\
\end{array}
\right.
\end{array}
\nonumber
\end{eqnarray}
\end{table}

\section{Numerical analysis and discussions}

In this section we carry out a global fit for $B \to PP$ using available experimental data to determine the $SU(3)/U(3)$ invariant amplitudes. In the numerical analysis we use Cabbibo-Kobayashi-Maskawa (CKM) parameters determined from other global analysis. We summarize the Wolfenstein parameters 
which determine CKM matrix elements in the following~\cite{pdg}
\begin{eqnarray}
\notag \lambda = 0.22543 \pm 0.00094 \,,\;\;A = 0.802 \pm 0.029 \,,\;\;\rho = 0.154\pm 0.0124\,,\;\;\eta =0.363\pm 0.0078\,.
 \end{eqnarray}
For experimental inputs of the branching ratios and $CP$ violating asymmetries,
we use the data in Refs.~\cite{pdg,average}, while for 
${\cal B}(B_d \to \pi^0 \eta)$ and ${\cal B}(B_s \to \eta' \eta')$ we use the newly observed ones
 from Refs.~\cite{Pal:2015ewa,Aaij:2015qga}, respectively.

To understand the significance of each type of amplitudes in explaining the data, 
we consider several different ways to carry out our numerical analysis. To see if indeed the annihilation contributions are smaller than non-annihilation amplitudes, we analyze the data in two different ways: with or without annihilation contributions. The analysis with or without $\eta$ and/or $\eta'$ in the final states may also be significantly different because the mixing effect of $\eta-\eta'$ may complicate the situation. We therefore also carry out analysis according to whether or not
include $\eta$ and/or $\eta'$ in the final states.  In the case with $\eta$ and/or $\eta'$ in the final states, by fitting data, one may also obtain some information about the mixing angle $\theta$. This may provide another way to determine the mixing angle.\\[2mm]
Our results are presented for four different cases:

1). Analysis without annihilation contributions and 
without $\eta$ and/or $\eta'$ in the final states. 

2). Analysis with annihilation contributions and 
without $\eta$ and/or $\eta'$ in the final states.  

3). Analysis without annihilation contributions and 
with $\eta$ and/or $\eta'$ in the final states. 

4). Analysis with annihilation contributions and 
with $\eta$ and/or $\eta'$ in the final states.\\

The values of the minimal $\chi^2$ per degrees of freedom (d.o.f) for different cases from our fit are given by
\begin{eqnarray}
Case\;\; 1), \;\;1.65\;;\;\;\;\;Case\;\; 2),\;\;  1.27\;;\;\;\;\; Case\;\; 3),\;\;1.71\;;\;\;\;\;Case\;\; 4),\;\;1.67.
\end{eqnarray}
Note that for each case, the minimal $\chi^2$ is different because the available decay modes for data fitting for each case is different. The above minimal $\chi^2$ per d.o.f indicate that all the four fits are reasonable ones. 
%====================
{\footnotesize
\begin{table}[h!]
\caption{The best fit values and their 68\% C.L. ranges for the
hadronic parameters in the four cases. }\label{DD}
\begin{tabular}{|c||cc|cc|}
\hline
&\multicolumn{2}{c}{without $\eta$ and $\eta^\prime$}\vline
&\multicolumn{2}{c}{with $\eta$ and $\eta^\prime$}\vline\\
&
Case 1)&Case 2)&Case 3)&Case 4)\\
\hline
$C^P_{\bar 3}$
&$0.142 \pm 0.001$ &$0.141 \pm 0.001$&$0.145 \pm 0.002$&$0.142 \pm 0.001$\\
$C^T_{\bar 3}$
&$-0.188 \pm 0.017$&$-0.198 \pm 0.026$&$-0.197 \pm 0.018$&$-0.211 \pm 0.027$\\
$C^T_6$
&$0.259 \pm 0.021$&$0.257 \pm 0.025$&$0.245 \pm 0.016$&$0.255 \pm 0.021$\\
$C^T_{\overline{15}}$
&$ -0.143 \pm 0.004$&$-0.141 \pm 0.004$&$-0.144 \pm 0.004$&$-0.142 \pm 0.004$\\
$\delta_{  \bar 3}$
&$(-121 \pm 5)^\circ$&$(-135 \pm 6)^\circ$&$(-124 \pm 5)^\circ$&$(-140 \pm 6)^\circ$\\
$\delta_6$
&$(50 \pm 4)^\circ$&$(54 \pm 6)^\circ$&$(51 \pm 4)^\circ$&$(56 \pm 6)^\circ$\\
$\delta_{\overline{15}}$
&$(169 \pm 4)^\circ$&$(171 \pm 4)^\circ$&$(165 \pm 3)^\circ$&$(172 \pm 3)^\circ$\\\hline
$A^T_{\bar 3}$
&---&$-0.034 \pm 0.015$&---&$-0.039 \pm 0.014$\\
$A^P_{\bar 3}$
&---&$-0.013 \pm 0.002$&---&$-0.013 \pm 0.002$\\
$A^T_{\overline{15}}$
&---&$-0.025 \pm 0.012$&---&$-0.020 \pm 0.012$\\
$\delta_{A^T_{\bar 3}}$
&---&$(-23 \pm 29)^\circ$&---&$(-16 \pm 25)^\circ$\\
$\delta_{A^P_{\bar 3}}$
&---&$(-120 \pm 16)^\circ$&---&$(-123 \pm 16)^\circ$\\
$\delta_{A^T_{\overline{15}}}$
&---&$(-30 \pm 26)^\circ$&---&$(-14\pm 27)^\circ$\\\hline\hline
$D^P_{\bar 3}$
&---&---&$-0.077 \pm 0.007$&$-0.073 \pm 0.008$\\
$D^T_{\bar 3}$
&---&---&$0.272 \pm 0.036$&$0.275 \pm 0.053$\\
$\delta_{D^P_{\bar 3}}$
&---&---   &$(-55 \pm 9)^\circ$  &$(-55 \pm 10)^\circ$\\
$\delta_{D^T_{\bar 3}}$
&---&---  &$(-90 \pm 9)^\circ$&$(-92 \pm 9)^\circ$\\\hline
$B^T_{\bar 6}$
&---&---&---&$0.099 \pm 0.094$\\
$B^T_{\overline{15}}$
&---&---&---&$-0.038 \pm 0.016$\\
$\delta_{B^T_{\bar 6}}$
&---&---   &---&$(75 \pm 55)^\circ$\\
$\delta_{B^T_{\overline{15}}}$
&---&---  &---&$(78 \pm 48)^\circ$\\\hline
$\theta$
&---&---  &$(-18.4\pm 1.2)^\circ$&$(-18.8\pm 1.2)^\circ$\\\hline
$\chi^2/d.o.f.$
&$1.65$&$1.27$  &$1.71$&$1.67$\\\hline
\end{tabular}
\end{table}
}
%====================

The hadronic parameters determined for the four cases mentioned above are listed in Table~\ref{DD}.
After the hadronic parameters are determined, one can predict some of  
the not-yet-observed branching ratios and CP violating asymmetries. The results are 
given in Tables~\ref{table4},\ref{table5},\ref{table6},\ref{table7}. In the following we comment on some features of our analysis.

As mentioned before,
the annihilation contributions $A_i$ are expected to be small compared with those of non-annihilation contributions $C_i$. Our fitting supports this expectation. The conclusions are drawn from comparing Case~1) with Case~2), and Case~3) with Case~4). 
Case~1) is 
an 
$SU(3)$ analysis neglecting annihilation contributions. A complete $SU(3)$ analysis would involve $\eta_8$. However, due to $\eta-\eta'$ mixing, one cannot obtain a complete information when $\eta_1$ is not included. But if one restricts the analysis to only include pions and Kaons in the final state, the analysis should give a reasonable fit if the annihilation contributions are indeed small. This is indeed supported by the smallness of the branching ratios for those 
%decay 
decays that
only receive annihilation contributions, such as $B_d\to K^- K^+$, $B_s\to \pi^+\pi^-$, and $B_s\to \pi^0\pi^0$. These modes only have
branching ratios of order $10^{-7}$. Analysis of Case 2) then helps to quantify the statement and obtain values for the relevant annihilation amplitudes. One can see that the annihilation amplitudes $A_i$ are several times smaller than the non-annihilation amplitudes $C_i$. The comparison of Case 3) with Case 4) also supports this conclusion. From Table~\ref{DD}, one can see that the current data still 
%left
leave the amplitude $D_i$ and $B_i$ with large errors. 
We hope that when more data become available, the $D_i$ and $B_i$ amplitudes will have better accuracy and the expectation that annihilation contributions are smaller than non-annihilation contributions will be tested further in the sector involving $\eta$ and $\eta'$ in $B \to PP$ decays. 

In case 3), there are 35 data points available with minimal $\chi^2/d.o.f$ of 1.71.
The LHCb has measured many more decay modes compared with that could be achieved 
by using data from Belle and Babar detectors at $B$-factories only.
In this case analysis with $\eta$ and $\eta'$ in the final states can be meaningfully carried out.
One can even obtain information about the $\eta-\eta'$ mixing angle. 
The $\eta-\eta'$ mixing angle determined from Case 3) analysis gives $\theta=(-18.4 \pm 1.2)^\circ$.
This is consistent with the value of $(-18\pm 2)^\circ$ 
%that 
given by Particle Data Group~\cite{pdg}.

Currently, the branching ratios and CP asymmetries for many decay modes 
with $\eta$ and $\eta'$ in the final states have not been observed, such as 
${\cal B}(B_d\to \eta\eta,\eta\eta',\eta'\eta')$ and ${\cal B}(B_s\to \eta\eta,\eta\eta')$.
Therefore, the theoretical predictions can be useful. 
For Case 3), the new parameters needed are $D_i$. 
The values for them are given in Table~\ref{DD}.
With the fitted $D_i$, we obtain
${\cal B}(B_u \to K^- \eta')$ and ${\cal B}(B_d \to \bar K^0 \eta')$
to be $(75.0^{+2.3}_{-2.7},65.0^{+2.7}_{-2.5})\times 10^{-6}$ 
%in 
which are
consistent with data. 
We note that ${\cal B}(B_s\to \eta\eta')$ around $24\times 10^{-6}$
can be as large as the observed ${\cal B}(B_s\to\eta'\eta')=(33\pm 11)\times 10^{-6}$,
while ${\cal B}(B_d\to \eta\eta,\eta'\eta')$ of order $10^{-7}$ agree with the experimental upper bounds.
When more data become available, this can be settled with high confidence.

In case 4),  the parameters $B_i$ with their phases, in principle, should be introduced implying 
8 new parameters. We find that the determinations of $B_3^T e^{i \delta_{B_3^T}}$ and $B_3^P e^{i \delta_{B_3^P}}$
require at least 4 data points from
$B_{d,s}\to \eta\eta,\eta\eta', \eta'\eta'$ decay modes,
but only ${\cal B}(B_s\to \eta'\eta')$ is available. Present available data cannot determine $B_3^T e^{i \delta_{B_3^T}}$ and $B_3^P e^{i \delta_{B_3^P}}$. Since they are annihilation amplitudes which are expected
to be small, we hence neglect their contributions for the practical fitting.  Therefore, in this case we will have 21 parameters to fit 
36 available data points. We obtain minimal $\chi^2/d.o.f$ to be 1.67 representing a reasonable fit.
Again in this case, we can determine the $\eta-\eta'$ mixing angle $\theta$ with
$\theta=(-18.8 \pm 1.2)^\circ$ represented to be stable compared to that in case 3).
The fitted $B_i$ have larger uncertainties, such as $B^T_{\bar 6}=0.099\pm 0.094$.
This is because that the data are not sufficient for the decays with $\eta_1$,
while $A_i$, $B_i$, $C_i$, and $D_i$ are fitted together.   
When more data become available, the predictions made for this case can be 
tests; in particular, data will tell whether the omission
of $B_3^T e^{i \delta_{B_3^T}}$ and $B_3^P e^{i \delta_{B_3^P}}$ for the fit is reasonable.

We now comment on a class of CP violating relations in the framework of $SU(3)$ flavor symmetry.  
This class of relations concern
the rate difference among some $B$ decays defined by
\begin{eqnarray}
\Delta(B\to PP)=\Gamma(B\to PP)-\Gamma({\bar B} \to \bar{P} \bar{P})\,,
\end{eqnarray}
which connects the branching ratio and the $CP$ violating asymmetry with
$\Delta(B_i \to PP)={\cal A}_{CP}(B_i \to PP){\cal B}(B_i \to PP)/\tau_{B_i}$
with $\tau_{B_i}$ the $B_i$ lifetime.

The unique feature of the SM in the CKM matrix elements that 
$Im(V_{ub} V^*_{ud} V^*_{tb} V_{td})=-Im(V_{ub} V^*_{us} V^*_{tb} V_{ts})$
can be used to relate the $\Delta S=0$ and $\Delta S=-1$ decay modes with the 
same tree amplitude $T$ and penguin amplitude $P$ which can be read off from Table~\ref{su3tb}. 
For instance, for  $B_s\to K^+ \pi^-$ and $B_d\to \pi^+K^-$, we obtain
\begin{eqnarray}
\frac{{\cal A}_{CP}({B_d}\to{\pi^+K^-})}{{\cal A}_{CP}({B_s}\to{K^+ \pi^-})}
+{\cal R}(\Delta^{B_d}_{\pi^+K^-}/\Delta^{B_s}_{K^+ \pi^-})
\frac{{\cal B}({B_s}\to{K^+ \pi^-})/\tau_{B_s}}{{\cal B}({B_d}\to{\pi^+K^-})/\tau_{B_d}}=0\,,
\end{eqnarray}
with ${\cal R}(\Delta^{B_d}_{\pi^+K^-}/\Delta^{B_s}_{K^+ \pi^-})=1$.

If annihilation amplitudes are neglected, there are additional relations, for example
\begin{eqnarray}
\frac{{\cal A}_{CP}({B_d}\to{\pi^+ K^-})}{{\cal A}_{CP}({B_d}\to{\pi^+\pi^-})}+
{\cal R}^{\prime}(\Delta^{B_d}_{\pi^+ K^-}/\Delta^{B_d}_{\pi^+\pi^-})
\frac{{\cal B}({B_d}\to{\pi^+\pi^-})}{{\cal B}({B_d}\to{\pi^+ K^-})}\simeq 0\,,
\end{eqnarray}
with ${\cal R}^{\prime}(\Delta^{B_d}_{\pi^+ K^-}/\Delta^{B_d}_{\pi^+\pi^-})\simeq 1$. 

Deviation of ${\cal R}_i$ away from 1 is a 
measure of $SU(3)$ flavor symmetry breaking. In Table \ref{Ri} we list ${\cal R}_i$ and ${\cal R}_i^\prime$ for some relations predicted with annihilation amplitudes and with annihilation amplitudes neglected, respectively. QCD based perturbation theory also predict similar values~\cite{He:2013vta,QCDF,Cheng:2005bg}.
Note that experimentally, 
${\cal R}_{data}(\Delta^{B_d}_{\pi^+K^-}/\Delta^{B_s}_{K^+ \pi^-})=1.12 \pm 0.22$ and 
${\cal R}^{\prime}_{data}(\Delta^{B_d}_{\pi^+ K^-}/\Delta^{B_d}_{\pi^+\pi^-})\simeq 1.02 \pm 0.19$. The $SU(3)$ predictions are in good agreement with data. Since the relation with annihilation contributions neglected is also in good agreement with data, this also provides an evidence that annihilation contributions are indeed small. If $SU(3)$ is exact the fitted central value for ${\cal R}_i$ should be equal to 1. The deviation in Table \ref{Ri} is due to the fact that in calculating the values, we have used physics Kaon and pion masses, branching ratios from fit and also experimental values for the lifetimes which slightly breaks $SU(3)$ flavor symmetry. Theoretically there are also several other pairs obeying
the relations discussed (listed in Table \ref{Ri}), at this moment there are large error bars to draw any conclusion. But once relevant quantities are measured, they will further test the theory.

In Table \ref{table7}, we notice that several CP asymmetries are determined to be large. This is because accidental cancellations in the amplitudes for relevant decays (large final state interaction phase) and need to be tested. This may also reflect the fact that data are not sufficient to constrain the amplitudes with high precision and the ``best'' fits are are some very shallow local minimuns. More data are required to draw meaningful conclusions.

Finally, we make a comment on the recent theoretical study in Ref.~\cite{HY_SU3} based on
the diagrammatic $SU(3)$ flavor symmetry. 
Our fittings include the newly observed 
${\cal B}(B_d \to \pi^0 \eta)$ and ${\cal B}(B_s \to \eta' \eta')$.
Despite of the measured ${\cal B}(B_d\to \eta\eta')<1.2\times 10^{-6}$, 
we predict ${\cal B}(B_d\to \eta\eta')$ to be $2\times 10^{-6}$
similar to that in Ref.~\cite{HY_SU3}. 
There is some tension between the fitted 
${\cal B}(B_d \to \pi^0 \eta)=(0.91\pm 0.03)\times 10^{-6}$ and 
the value of $(0.41 \pm 0.22)\times 10^{-6}$ from the data in comparison with
${\cal B}(B_d \to \pi^0 \eta)=(0.12\pm 0.07)\times 10^{-6}$~\cite{HY_SU3}.
Note that the predictions for ${\cal B}(B_d \to \pi^0 \eta)$ in the approaches of 
QCD factorization, pQCD, and SCET~\cite{QCDF,pQCD,SCET}
are of order $10^{-8}$. Future experiments can provide information to test these predictions.

%==================

\section{conclusions}

In this work we have performed an updated global analysis for $B \to PP$ using the latest experimental data 
based on flavor symmetry. Without including $\eta$ and $\eta'$ in the final states, $SU(3)$ flavor symmetry is sufficient for the analysis.
In order to include $P$ being $\eta$ or $\eta'$ in the analysis, we enlarged the symmetry to 
$U(3)$ flavor symmetry. In this case we also took into account  $\eta-\eta'$ mixing effect. 
We found that $SU(3)$ flavor symmetry can explain data well without $P$ being $\eta$ or $\eta'$. 

We have considered four different scenarios for data fitting to see how annihilation and also how inclusion of $\eta$ and $\eta'$ affect the results. The annihilation amplitudes 
were found to be small consistent with expectations. Current available data could
give constraints on the amplitudes which induce $P = \eta,\;\eta'$ decays in the framework of $U(3)$ flavor symmetry. The $\eta-\eta'$ mixing angle $\theta$ could
also be determined with $\theta = (-18.4\pm1.2)^\circ$ which is consistent with the value given by Particle Data Group from other fittings~\cite{pdg}. Several CP violating relations predicted by $SU(3)$ flavor symmetry were
found in good agreement with data. Although current data 
could not fix two annihilation amplitudes $B^{T,P}_3 e^{i\delta_{B_3^{T,P}}}$, as they 
were expected to be small, we were
able to predict several $B \to PP$ decay modes which have not been measured. 
These predicted branching ratios are accessible at the LHCb. We look forward to
more data to come  to test the framework of $SU(3)/U(3)$ flavor symmetry for $B$ decays.

\begin{acknowledgments}

The work was supported in part by MOE Academic Excellent Program (Grant No: 102R891505) and MOST of ROC, and in part by NSFC(Grant No:11175115 and 11575111) and Shanghai Science and Technology Commission (Grant No: 11DZ2260700) of PRC. We thank Tom Browder for bringing Ref.\cite{belle-new} to our attention.

\end{acknowledgments}

\noindent
\\

Note Added: After submitting our paper to arXiv, we became aware that $B_s \to K^0\bar K^0$ has been measured 
recently by Belle collaboration~\cite{belle-new} with a branching ratio of 
$(19.6^{+5.8}_{-5.1}(stat.)\pm 1.0(sys.)\pm 2.0 (N_{B^0_s\bar B^0_s}))\times 10^{-6}$. The measured branching ratio in good agreement with our prediction.

\newpage
%==================
{\footnotesize
\begin{table}[htb]
\caption{The central values and 68\% C.L. allowed ranges for
branching ratios (in units of $10^{-6}$), 
where the superscript $a$ denotes that the decay without $C_i$
is not involved in the fitting.}\label{table4}
\begin{tabular}{|c||c|cccc|}
\hline
\bf Branching ratios&data&case1&case2&case3&case4\\\hline
%1
$B_u \to \pi^- \pi^0$
&$5.48 \pm 0.35$
&$5.57^{+ 0.14}_{-0.13}$&$5.42^{+ 0.14}_{-0.13}$  &$5.69^{+ 0.13}_{-0.13}$&$ 5.54^{+ 0.13}_{-0.12}$\\
%2
$B_u \to K^- K^0$
&$ 1.32 \pm 0.14 $
&$ 1.34^{+ 0.04}_{-0.04} $&$1.34^{+ 0.08}_{-0.06} $&$ 1.20^{+ 0.04}_{-0.03} $&$ 1.18^{+ 0.07}_{-0.05} $\\
%3
$B_d \to \pi^+ \pi^-$
&$5.10\pm 0.19$
&$5.20^{+ 0.14}_{-0.14}$&$5.12^{+ 0.22}_{-0.20}$&$5.22^{+ 0.14}_{-0.13}$&$5.13^{+ 0.23}_{-0.20}$\\
%4
$B_d \to\pi^0 \pi^0$
&$1.17 \pm 0.13$
&$1.05^{+ 0.04}_{-0.04}$&$1.15^{+ 0.06}_{-0.05}$&$1.06^{+ 0.04}_{-0.03}$&$1.17^{+ 0.05}_{-0.05}$\\
%5
$B_d \to \bar K^0 K^0$
&$1.21 \pm 0.16$
&$1.23^{+ 0.04}_{-0.03}$&$1.31^{+ 0.07}_{-0.05}$&$1.10^{+ 0.03}_{-0.03}$&$1.31^{+ 0.08}_{-0.06}$\\
%6
$B_u \to \pi^- \bar K^0$
&$  23.79 \pm 0.75 $
&$ 23.18^{+ 0.13}_{-0.13}$&$ 22.72^{+ 0.15}_{-0.14}$&$ 23.05^{+ 0.12}_{-0.12}$&$ 22.73^{+ 0.14}_{-0.14}$\\
%7
$B_u \to \pi^0 K^-$
&$12.94 \pm 0.52$
&$13.03^{+ 0.08}_{-0.08}$&$12.78^{+ 0.08}_{-0.08}$&$13.00^{+ 0.08}_{-0.08}$&$12.83^{+ 0.08}_{-0.08}$\\
%8
$B_d \to \pi^+ K^-$
&$19.57 \pm 0.53$
&$20.64^{+ 0.12}_{-0.12}$&$20.60^{+ 0.14}_{-0.13}$&$20.84^{+ 0.12}_{-0.12}$&$20.72^{+ 0.13}_{-0.12}$\\
%9
$B_d \to \pi^0 \bar K^0$
&$9.93 \pm 0.49$
&$9.20^{+0.06}_{-0.06}$&$9.15^{+0.06}_{-0.06}$&$9.28^{+0.06}_{-0.06}$&$9.20^{+0.06}_{-0.06}$\\
%10
$B_d \to K^+ K^-$
&$0.13 \pm 0.05$
&-----$^a$&$0.14^{+0.03}_{-0.02}$&-----$^a$&$0.14^{+0.03}_{-0.02}$\\
\hline
%s1
$B_s \to K^+ \pi^-$
&$5.5 \pm 0.5$
&$5.0^{+0.1}_{-0.1}$&$5.57^{+0.19}_{-0.19}$&$5.01^{+0.13}_{-0.13}$&$5.61^{+0.20}_{-0.17}$\\
%s2
$B_s \to K^0 \pi^0$
&-----
&$2.02^{+0.08}_{-0.07}$&$1.59^{+0.08}_{-0.07}$&$2.04^{+0.07}_{-0.07}$&$1.64^{+0.08}_{-0.06}$\\
%s3
$B_s \to K^+K^- $
&$24.8\pm 1.7$
&$19.8^{+0.1}_{-0.1}$&$24.5^{+0.6}_{-0.6}$&$20.0^{+0.1}_{-0.1}$&$24.5^{+0.6}_{-0.6}$\\
%s4
$B_s \to K^0 \bar K^0$
&$  <66$
&$20.5^{+0.1}_{-0.1}$&$22.9^{+0.3}_{-0.3}$&$20.4^{+0.1}_{-0.1}$&$22.4^{+0.4}_{-0.3}$\\
%s5
$B_s \to \pi^+ \pi^-$
&$  0.76 \pm 0.19$
&-----$^a$&$0.72^{+0.06}_{-0.05}$&-----$^a$&$0.71^{+0.06}_{-0.05}$\\
%s6
$B_s \to \pi^0 \pi^0$
&$  <210$
&-----$^a$&$0.18^{+0.01}_{-0.01}$&-----$^a$&$0.18^{+0.01}_{-0.01}$\\
\hline
\end{tabular}
\end{table}
}
%==================
{\footnotesize
\begin{table}[htb]
\caption{The central values and 68\% C.L. allowed ranges for
CP asymmetries (in units of $10^{-2}$).}\label{table5}
\begin{tabular}{|c||c|cccc|}
\hline
%&\multicolumn{4}{c}{\bf Branching Ratios}\vline\\
%&\multicolumn{2}{c}{\bf CP Asymmetries}\vline \\%\hline
\bf CP asymmetries&data&case1&case2&case3&case4\\\hline
%1
$B_u \to \pi^- \pi^0$
&$2.6\pm 3.9$
&$0 \pm 0$&$0 \pm 0$  &$0 \pm 0$&$0 \pm 0$\\
%2
$B_u \to K^- K^0$
&$-8.7\pm 10.0$
&$ -2.8^{+ 4.0}_{-4.0} $&$-3.8^{+7.4}_{-6.8}$&$ -5.5^{+ 3.8}_{-3.8} $&$ -7.7^{+ 8.6}_{-7.2} $\\
%3
$B_d \to \pi^+ \pi^-$
&$31 \pm 5$
&$31.1^{+ 1.1}_{-1.1}$&$30.2^{+2.2}_{-2.4}$&$31.1^{+ 1.1}_{-1.1}$&$29.7^{+ 2.0}_{-2.1}$\\
%4
$B_d \to\pi^0 \pi^0$
&$43.0  \pm 24.0$
&$57.2^{+ 1.2}_{-1.3}$&$64.0^{+ 1.8}_{-1.9}$&$56.1^{+ 1.2}_{-1.2}$&$63.3^{+ 1.7}_{-1.8}$\\
%5
$B_d \to \bar K^0 K^0$
&$-60.0\pm 70.0$
&$-2.8^{+ 4.0}_{-4.0}$&$-17.8^{+ 9.7}_{-8.6}$&$-5.5^{+ 4.0}_{-3.8}$&$-18.0^{+ 9.2}_{-8.1}$\\
%6
$B_u \to \pi^- \bar K^0$
&$-1.7\pm 1.6$
&$ 0.17^{+ 0.24}_{-0.24}$&$ 0.23^{+ 0.64}_{-0.47}$&$ 0.30^{+ 0.22}_{-0.22}$&$ 0.42^{+ 0.39}_{-0.48}$\\
%7
$B_u \to \pi^0 K^-$
&$4.0\pm 2.1$
&$5.8^{+ 0.5}_{-0.5}$&$4.2^{+ 0.4}_{-0.7}$&$5.8^{+ 0.4}_{-0.4}$&$4.7^{+ 0.6}_{-0.6}$\\
%8
$B_d \to \pi^+ K^-$
&$-8.2   \pm 0.6$
&$-7.8^{+ 0.3}_{-0.3}$&$-8.1^{+ 0.4}_{-0.4}$&$-7.9^{+ 0.3}_{-0.3}$&$-8.0^{+ 0.4}_{-0.4}$\\
%9
$B_d \to \pi^0 \bar K^0$
&$0\pm 13$
&$-13.3^{+0.5}_{-0.5}$&$-11.3^{+0.5}_{-0.5}$&$-13.2^{+0.4}_{-0.4}$&$-11.6^{+0.5}_{-0.5}$\\
%10
$B_d \to K^+ K^-$
&-----
&-----$^a$&$82.8^{+4.4}_{-6.0}$&-----$^a$&$83.6^{+4.4}_{-6.2}$\\
\hline
%s1
$B_s \to K^+ \pi^-$
&$26.0\pm4.0$
&$31.1^{+1.1}_{-1.2}$&$28.1^{+1.4}_{-1.3}$&$31.2^{+1.0}_{-0.1}$&$28.0^{+1.2}_{-1.1}$\\
%s2
$B_s \to K^0 \pi^0$
&-----
&$57.2^{+1.2}_{-1.3}$&$61.4^{+1.8}_{-2.1}$&$55.9^{+1.2}_{-1.2}$&$60.6^{+1.6}_{-1.9}$\\
%s3
$B_s \to K^+K^- $
&$-14\pm 11$
&$-8.0^{+0.3}_{-0.4}$&$-5.6^{+0.5}_{-0.5}$&$-8.0^{+0.3}_{-0.3}$&$-5.6^{+0.4}_{-0.5}$\\
%s4
$B_s \to K^0 \bar K^0$
&-----
&$0.17^{+0.24}_{-0.23}$&$12.1^{+1.2}_{-1.3}$&$0.27^{+0.21}_{-0.22}$&$10.4^{+1.3}_{-1.4}$\\
%s5
$B_s \to \pi^+ \pi^-$
&-----
&-----$^a$&$-16.1^{+1.9}_{-1.6}$&-----$^a$&$-16.2^{+2.1}_{-2.1}$\\
%s6
$B_s \to \pi^0 \pi^0$
&$  <210$
&-----$^a$&$-16.1^{+1.9}_{-1.9}$&-----$^a$&$-16.2^{+2.0}_{-2.0}$\\
\hline
\end{tabular}
\end{table}
}
%==================
{\footnotesize
\begin{table}[h]
\caption{The central values and their 68\% C.L. allowed ranges for
branching ratios (in units of $10^{-6}$) with
at least one of the final mesons to be a $\eta$ or $\eta'$.}\label{br2} \label{table6}
\begin{tabular}{|c||c|cc|}
\hline
%&\multicolumn{2}{c}{\bf Branching Ratios}\vline
%&\multicolumn{2}{c}{\bf CP Asymmetries}\vline \\%\hline
{\bf Branching ratios}&data&case3&case4\\\hline
$B_u \to \pi^- \eta$
&$4.02 \pm 0.27$
&$3.77^{+0.12}_{-0.11}$&$3.73^{+1.50}_{-0.45}$\\
$B_u \to \pi^- \eta'$
&$  2.7 \pm 0.5 $
&$ 3.33^{+0.19}_{-0.16}$&$ 3.23^{+8.81}_{-0.92}$\\
$B_d \to \pi^0 \eta$
&$ 0.41 \pm 0.22 $
&$ 0.91^{+0.03}_{-0.03}  $&$ 0.77^{+0.61}_{-0.02}  $\\
$B_d \to \pi^0 \eta'$
&$  1.2 \pm 0.4 $
&$ 1.06^{+0.06}_{-0.05}  $&$ 1.23^{+4.21}_{-0.11}  $\\
$B_u \to K^- \eta$
&$  2.36 \pm 0.22 $
&$ 2.16^{+0.22}_{-0.17}  $&$ 2.19^{+0.37}_{-0.24}  $\\
$B_u \to K^- \eta'$
&$  71.1 \pm 2.6 $
&$ 75.0^{+2.3}_{-2.7}  $&$ 71.1^{+4.7}_{-3.6}  $\\
$B_d \to \bar K^0 \eta$
&$  1.23 \pm 0.27 $
&$ 1.63^{+0.19}_{-0.15}  $&$ 1.54^{+0.28}_{-0.17}  $\\
$B_d \to \bar K^0 \eta'$
&$  66.1 \pm 3.1 $
&$ 65.0^{+2.7}_{-2.5}  $&$ 64.5^{+4.2}_{-3.4}  $\\
$B_d \to \eta \eta$
&$<1.0$
&$ 0.33^{+0.02}_{-0.01}  $&$0.55^{+0.84}_{-0.11}$\\
$B_d \to \eta \eta'$
&$ <1.2$
&$ 1.91^{+0.10}_{-0.10}  $&$3.33^{+10.06}_{-0.66}$\\
$B_d \to \eta' \eta'$
&$<1.7$
&$ 0.41^{+0.03}_{-0.02}  $&$0.28^{+0.92}_{-0.02}$\\
\hline
$B_s\to K \eta$
&-----
&$ 0.99^{+0.04}_{-0.04}  $&$ 1.12^{+1.84}_{-0.29}  $\\
$B_s \to K\eta'$
&-----
&$ 3.52^{+0.16}_{-0.14}  $&$ 4.29^{+10.29}_{-0.48}  $\\
$B_s \to \pi^0 \eta$
&$  <1000 $
&$ 0.048^{+0.002}_{-0.002}  $&$ 0.037^{+0.13}_{-0.01}  $\\
$B_s \to \pi^0 \eta'$
&-----
&$ 0.085^{+0.003}_{-0.003}  $&$ 0.25^{+1.24}_{-0.06}  $\\
$B_s \to \eta \eta$
&$  <1500 $
&$ 2.81^{+0.12}_{-0.11}  $&$3.29^{+0.13}_{-0.06}$\\
$B_s \to \eta \eta'$
&-----
&$ 23.70^{+0.65}_{-0.54}  $&$21.99^{+0.58}_{-0.13}$\\
$B_s \to \eta' \eta'$
&$  33.1 \pm 10.4 $
&$ 21.30^{+1.10}_{-0.90}  $&$20.42^{+1.15}_{-1.00}$\\
\hline
\end{tabular}
\end{table}
}
%%%%%%%%%%%%%%%%%%%%%%%%%%%%%%%%%%%%%%%%%
%==================
{\footnotesize
\begin{table}[h]
\caption{The central values and their 68\% C.L. allowed ranges for
CP asymmetries (in units of $10^{-2}$) with
at least one of the final mesons to be a $\eta$ or $\eta'$.}\label{br2} \label{table7}
\begin{tabular}{|c||c|cc|}
\hline
{\bf CP asymmetries}&data&case3&case4\\\hline
$B_u \to \pi^- \eta$
&$-14   \pm 5$
&$-14.6^{+2.8}_{-2.7}$&$-12.3^{+28.5}_{-20.9}$\\
$B_u \to \pi^- \eta'$
&$6  \pm 15$&$8.9^{+5.9}_{-6.3}$&$5.6^{+22.8}_{-23.4}$\\
$B_d \to \pi^0 \eta$
&-----&$ -26.8^{+4.2}_{-3.9}$&$-0.4^{+30.4}_{-26.7}$\\
$B_d \to \pi^0 \eta'$
&-----&$-48.5^{+7.6}_{-6.5}$&$83.3^{+5.2}_{-57.6}$\\ %64.6 isn't type
$B_u \to K^- \eta$
&$-37 \pm 8$&$-30.9^{+2.3}_{-2.4}$&$-31.1^{+13.3}_{-9.9}$\\
$B_u \to K^- \eta'$
&$1.3   \pm 1.7$&$0.5^{+0.3}_{-0.3}$&$0.8^{+6.8}_{-7.5}$\\
$B_d \to \bar K^0 \eta$
&-----&$3.2^{+1.8}_{-2.2}$&$8.7^{+16.8}_{-12.2}$\\
$B_d \to \bar K^0 \eta'$
&-----&$4.3^{+0.3}_{-0.3}$&$34.8^{+7.4}_{-6.9}$\\
$B_d \to \eta \eta$
&-----&$-86.6^{+2.0}_{-1.6}$&$-42.1^{+53.1}_{-2.6}$\\
$B_d \to \eta \eta'$
&-----&$-68.8^{+5.4}_{-4.3}$&$-27.9^{+51.9}_{-6.7}$\\
$B_d \to \eta' \eta'$
&-----&$-62.7^{+6.4}_{-5.5}$&$-87.9^{+56.5}_{-10.8}$\\
\hline
$B_s\to K \eta$
&-----&$-5.5^{+3.4}_{-3.4}$&$-11.5^{+28.8}_{-13.4}$\\
$B_s \to K\eta'$
&-----&$-79.7^{+4.1}_{-3.1}$&$-93.0^{+62.6}_{-2.1}$\\
$B_s \to \pi^0 \eta$
&-----&$98.1^{+0.4}_{-0.7}$&$83.3^{+4.8}_{-57.3}$\\
$B_s \to \pi^0 \eta'$
&-----&$98.1^{+0.4}_{-0.7}$&$64.7^{+10.0}_{-35.4}$\\
$B_s \to \eta \eta$
&-----&$-13.5^{+0.4}_{-0.4}$&$6.0^{+2.1}_{-3.2}$\\
$B_s \to \eta \eta'$
&-----&$-3.1^{+0.3}_{-0.4}$&$-1.3^{+2.5}_{-1.3}$\\
$B_s \to \eta' \eta'$
&-----&$4.5^{+0.4}_{-0.4}$&$4.8^{+4.5}_{-3.7}$\\
\hline
\end{tabular}
\end{table}
}
%%%%%%%%%%%%%%%%%%%%%%%%%%%%%%%%%%%%%%%%%

\begin{table}[b]
\caption{${\cal R}^{(\prime)}_i$ to test the $SU(3)$ flavor symmetry.
The fitted numbers in the parentheses are for case 1) and case 2), respectively.}\label{Ri}
\begin{tabular}{|c|cc|}
\hline
modes&${\cal R}_{data}$&${\cal R}^{(\prime)}_{fit}$\\
\hline
${\cal R}(\Delta^{B_d}_{\pi^+K^-}/\Delta^{B_s}_{K^+ \pi^-})$ %4
&$1.12\pm 0.22$&$( 1.03\pm 0.06 , 1.06\pm 0.08)$\\
${\cal R}(\Delta^{B_s}_{ K^+ K^-}/\Delta^{B_d}_{\pi^+\pi^-})$ %2
&$2.20\pm 1.77$&$(0.98\pm 0.06, 0.89\pm 0.12)$\\
${\cal R}(\Delta^{B_u}_{\pi^- \bar K^0}/\Delta^{B_u}_{K^- K^0})$ %1
&$-3.52\pm 5.25$&$(1.05 \pm 2.07, 1.02\pm 3.48)$\\
${\cal R}(\Delta^{B_d}_{\pi^0 \bar K^0}/\Delta^{B_s}_{K^0 \pi^0})$ %3
&---&$(1.06 \pm 0.06, 1.06\pm 0.08)$\\
${\cal R}(\Delta^{B_s}_{\pi^+ \pi^-}/\Delta^{B_d}_{K^- K^+})$ %5
&---&(---, $1.00\pm 0.27)$\\
${\cal R}(\Delta^{B_s}_{\pi^0 \pi^0}/\Delta^{B_d}_{K^- K^+})$ %6
&---&(---, $1.00 \pm 0.02)$\\
\hline\hline
${\cal R}'(\Delta^{B_d}_{\pi^+ K^-}/\Delta^{B_d}_{\pi^+\pi^-})$
&$1.02\pm 0.19$&$( 0.99\pm 0.06, 1.07\pm 0.11)$\\
${\cal R}'(\Delta^{B_d}_{\pi^0 \bar K^0}/\Delta^{B_d}_{\pi^0 \pi^0})$
&$0.00\pm 1.28$&$(1.02 \pm 0.06, 0.70\pm 0.05)$\\
${\cal R}'(\Delta^{B_s}_{K^+K^-}/\Delta^{B_s}_{K^+ \pi^-})$ %4
&$2.42\pm 1.96$&$(1.01\pm 0.06, 0.88\pm 0.10)$\\
${\cal R}'(\Delta^{B_u}_{\pi^- \bar K^0}/\Delta^{B_d}_{\bar K^0 K^0})$ %3
&$-0.56\pm 0.83$&$(1.14 \pm 2.28, 0.22\pm 0.64)$\\
\hline
\end{tabular}
\end{table}

\end{document}